\documentclass[aps,pre,preprint,superscriptaddress,showpacs,floatfix]{revtex4-1}

\usepackage{epsfig}
\usepackage{latexsym}
\usepackage{amsmath}
\usepackage{amssymb}
\usepackage{amsfonts}
\usepackage{graphicx}
\usepackage{wrapfig}
\usepackage{url}

\begin{document}

\title{Universality class of balanced flows with bottlenecks: granular flows, pedestrian fluxes and financial price dynamics}

\author{Daniel R. Parisi}
\email[]{dparisi@itba.edu.ar}
\affiliation{Buenos Aires Institute of Technology, 25 de Mayo 444, (1002) C. A. de Buenos Aires, Argentina.}
\affiliation{National Council for Scientific and Technical Research (CONICET). Argentina.}

\author{Didier Sornette}
\email[]{dsornette@ethz.ch}
\affiliation{Department of Management, Technology and Economics, ETH Z\"{u}rich, Kreuzplatz 5, CH-8032 Z\"{u}rich CH-1015
Switzerland.}

\author{Dirk Helbing}
\email[]{dhelbing@ethz.ch}
\affiliation{Chair of Sociology, in particular of Modeling and Simulation, ETH Z\"{u}rich, Clausiusstrasse 50, CH-8092 Z\"{u}rich, Switzerland.}

\date{\today}



\begin{abstract}
We propose and document the evidence for an analogy between the dynamics of granular counter-flows
in the presence of bottlenecks or restrictions and financial price formation processes.
Using extensive simulations, we find that the counter-flows of simulated pedestrians through a door display many stylized facts observed in financial markets when the density around the door is compared with the logarithm of the price. The stylized properties are present already when the agents in the pedestrian model are assumed to display a zero-intelligent behavior. If agents are given decision-making capacity 
and adapt to partially follow the majority, periods of herding behavior may additionally occur. This generates the very slow decay of the autocorrelation of absolute return due to an intermittent dynamics. 
Our finding suggest that the stylized facts in the fluctuations of the financial 
prices result from a competition of two groups with opposite interests in the presence
of a constraint funneling the flow of transactions to a narrow band of prices.
\end{abstract}

\maketitle
\section{Introduction}
\label{intro}

Consider systems endowed with the following attributes:
\begin{enumerate}
\item[(a)] Competition in the form of opposite tendencies, such as
counter flows of particles moving in opposite directions, buyer-seller 
opposite drives to acquire versus drop an asset;
or liquidity providers/takers with opposite needs with respect cash and access to markets;
\item[(b)] Bottleneck or restriction/constriction that provide a convergent constraint
in the free flow in the system; in granular flows, this is in the form of a funnel 
or an opening separating two different spatial domains; in finance, this is 
associated with the fact that actual transactions occur in the limit of 
small or vanishing liquidity \cite{Bouchaudliquid}; In other words, 
whatever their volume, all orders
have to be funneled to a small price window in order to be executed.
\end{enumerate}
We conjecture that these systems operate in a ``pre-jammed'' state
with large intermittent fluctuations that exhibit
the following set of stylized facts:
\begin{enumerate}
\item Fat tail distribution of fluctuation amplitudes of some order parameter such as
density variations or log-price variations;
\item Tendency for the above distributions to converge slowly to the Gaussian
law at large space or time scales over which the order parameter fluctuations are measures;
\item weak and fast decaying auto-correlation of the signed fluctuations of the order parameter;
\item long-range auto-correlation of the amplitude (or ``volatility'') of the order parameter
fluctuations;
\item Hurst exponent and persistence in the dynamics of the volatility;
\item Scaling of the peaks of the distribution of the order parameter fluctuations;
\item Multifractality;
\item Existence of transient coherent regimes (bubbles, solitary waves, coherent structures)
bursting in crashes or fast and strong reorganization processes.
\end{enumerate}
This suggests a deep analogy
between the dynamics of granular counter-flows
in the presence of bottlenecks or restrictions and financial price formation processes.
The former applies to pedestrians in confined geometries or more
generally to flows of granular media in the presence of constrictions and constraints.

Analogies between complex flows and financial markets are not new.
In 1996, Ghashghaie et al. have shown that the distribution of velocity
increments of fully developed
turbulence and that of exchange rate fluctuations exhibit striking similarities \cite{Ghashghaie}. 
This led these authors to suggest a common connection via the existence
of cascades in both systems, Kolmogorov energy cascade in turbulence \cite{Frisch} and
information cascade in finance. However, Arneodo et al.  \cite{Arneodorebutcasc} remarked that
the two problems differ on the fundamental property of correlations and higher-order
statistics.  Indeed, spatial correlations in turbulence lead to the famous $-{5 \over 3}$ power-law for the
spectrum of the velocity fluctuations \cite{Frisch}, while no temporal 
correlations of the sort are visible in the power-spectrum of financial time series.
If such correlations existed in finance,  it would be
easy to use them to earn money, while the core of the problem in
turbulence is the existence of very strong correlations. The analogy between
turbulence and finance just based on a one-point statistics turned out to be a dead end.
Let us also mention the formalism of Vamo\c{s} et al. \cite{vamos2000} that counts
the flux of price changes in a universe of assets, which is similar to 
an hydrodynamic conservation equation.

Perhaps less fancy than hydrodynamic turbulence but more appropriate,
Bouchaud et al. \cite{molasseBouchaud} have suggested an analogy with 
molasses, the rock conglomerates that form as a result of geological sedimentary processes.
They proposed a model of financial fluctuations
based on the competition between liquidity providers and liquidity takers,
in which the existence of an excess flow of limit orders opposing the market order flow together with
a systematic anti-correlation of the bid-ask motion between trades
lead to create a `liquidity molasse' which dampens market volatility.

The present work extends these ideas to suggest the existence
of what could be referred to as a new ``universality class''
for out-of-equilibrium complex extended dynamical
systems characterized by ``balanced flows with bottlenecks''
endowed with the characteristics outlined above, and which are 
described by the set of properties 1-8. 
The suggestion of a similarity between pedestrian counter-flows throughout constrictions
and financial markets was first proposed by Helbing \cite{Helbing2001,Helbing2001a}. First supportive evidence of this idea was reported by Parisi \cite{Parisi2010}, considering the mean velocity near the door as the observable of the pedestrian system.
Our concept also extends models of the
 price formation process described via the order book dynamics of diffusing particles in 
 one dimension \cite{Bak1997,Svorencik2007}. It is compatible with the view
 that financial markets operate close to a critical point in a precise sense
 \cite{Bouchaudliquid,molasseBouchaud,VladDid}.
 
In the present work, we show that the dynamics of the density fluctuations 
at the location of the bottleneck reproduces many stylized facts already
documented for financial price fluctuations \cite{Daco,Cont2001,Bouchaud2001,Lux,Malsor}. 
We consider both the case of zero-intelligent pedestrians or particles that follow mechanical rules
and the extension where pedestrians can change their strategy
by imitating the majority. The later ingredient turns out to be necessary
to generate the equivalent of bubbles and crashes, while the
other stylized facts remain the same.

Our paper is organized as follows. In Section 2, we describe the simulated 
pedestrian system and the price time series used for comparison.
Section 3 presents the analysis of the eight statistical properties
mentioned above, obtained from simulations of the constrained
pedestrian system and from empirical financial data.
Section 4 presents our conclusions.

\section{Financial and Pedestrian Time Series}
\label{s2}
\subsection{Financial Data}
\label{s21}
Data from foreign exchange rates and stock indexes were analyzed in order to compare their stylized properties with those produced by a pedestrian simulation model.
Specifically, we evaluated the following data: EURUSD-1min; EURUSD-10min; EURUSD-1hour; CHFUSD-1hour; Nasdaq100-1hour; DJI-1day; NYSE100-10min, NYSE100-1hour and NYSE100-1day.
The financial time series contained between 10,000 and 30,000 data points and were taken from public internet sources \cite{DataSource1,DataSource2}. 
\subsection{Pedestrian Model and Setup}
\label{s22}

\subsubsection{Description of the model}

Our pedestrian simulations were based on the Social Force Model \cite{Helbing2000}. In this model, pedestrians are treated like circular discs with different radii representing the space occupied by them. The dynamics of each pedestrian ($i$) is governed by three forces: the ``driving force" ($\mathbf{F}_{Di}$), the ``contact force" ($\mathbf{F}_{Ci}$), and the ``social force" ($\mathbf{F}_{Si}$).
The driving force is responsible for the self-propulsion of each simulated pedestrian (`agent'), and it provides a constant input of energy into the system. The contact force is a dissipative and repulsive interaction force between particles that appears only when at least two particles overlap. The specifications of these two forces, and the parameter values were chosen as in Ref. \cite{Helbing2000}. However,  in the present case, the social force is assumed to be repulsive locally and attractive at a larger distance, as defined by Eqs. (\ref{F_S}) and (\ref{A}):
\begin{equation}
\mathbf{F}_{Si}=\sum_{j=1,j\neq i}^{N_{p}}~A(d _{ij})~\exp {\left(\frac{-d_{ij}}{B}\right)}~\mathbf{e}_{ij}^{n}~.
 \label{F_S}
\end{equation}
Here, $N_p$ is the total number of pedestrians in the system, $\mathbf{e}_{ij}^{n}$ is the unit vector pointing from pedestrian $j$ to $i$ (the `normal' direction), $d_{ij}$ is the edge-to-edge distance between both pedestrians (defined as the distance between their centers minus the sum of their radii, as in Ref. \cite{Helbing2000}), $B=0.08$ m is a constant determining the range of the social interaction, and $A$ is 
\begin{equation}
A(d _{ij})= \begin{cases}
+ 2000 \text{ N} & \text{ if $d_{ij}  <  0.15$ m  (repulsive)} \\
- 2000 \text{ N} & \text{ if $d_{ij}  >  0.15$ m  (attractive)} \\
\end{cases}
\label{A}
\end{equation}
where N stands for the unit of force in Newtons.

For the interaction of a pedestrian $i$ with a wall ($w$), the value of $A$ is assumed to be greater than zero for all $d_{iw}$ (i.e.: the wall has only repulsive effect).

The pedestriansÕ parameters, namely, mass ($m$), diameter ($2r$), and desired velocity ($v_d$) were uniformly distributed within the following ranges:
$m~\epsilon$~[70 kg,~90 kg], $2r~\epsilon$ [0.44 m,~0.56 m], and $v_d~\epsilon$ [1.05 m/s,~1.35 m/s], 
\subsubsection{The Simulated Pedestrian System}
\label{s221}
\begin{figure}
\begin{center}
\centerline{ \includegraphics[scale=0.5]{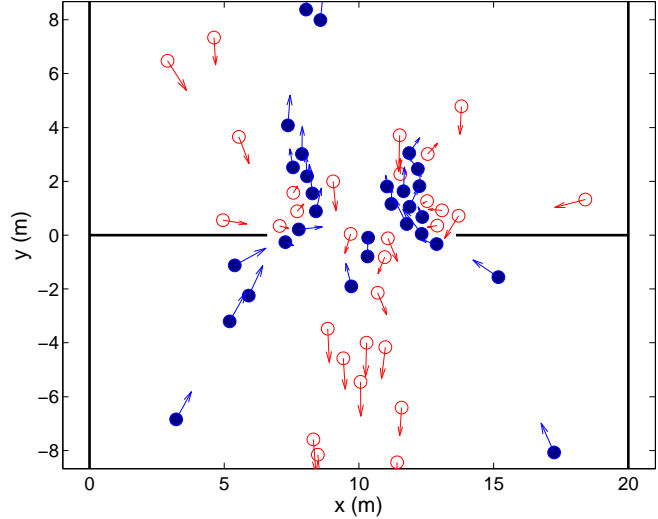}}
\caption{Setup of our simulation of pedestrian counterflows. The two kinds of particle filling indicate the state of the pedestrians. The arrows display their current velocities.}
\label{Fig_PCS}
\end{center}
\end{figure}

Our simulation of pedestrian counterflows is based on two open corridors of 20 m width, connected by a door of width $L$, as shown in Fig. \ref{Fig_PCS}. 
Initially, there are $N_p / 2$ agents on each side of the door. These agents attempt to reach the end of the corridor on the other side. In order to achieve this goal, the pedestrians must cross the door in opposite directions, and thus a counterflow is created.
Once a particle reaches a target placed 10 meters away from the door, it is instantaneously reinserted at a random position in the corridor where it started (not more than 10 meters from the door). In this way, a continuous counterflow is established with a fixed number of particles. This mechanism describes an automaton-like behavior of agents in the sense that they have a unique objective: to reach their assigned target, no matter where they are. We call this version of the counterflow system ``automata pedestrians" or ``automata agents".
\subsubsection{Agents with Decision-Making Capacity}
\label{s222}
In the pedestrian system described above, there are two groups of ``automata agents" with opposite, but fixed flow directions; let us call them $a$ and $b$. Because the number of agents belonging to each class is fixed ($N_a = N_b$), the system is forced to be in a statistical equilibrium from a population point of view. The model can be generalized by allowing agents to change their state via a decision mechanism. Then, the system is able to exhibit collective or herding behavior of the agents. For example, all agents could choose to be in the same state, i.e.,  to have the same desired walking direction. This situation is analogous to a financial crash, when all agents want to sell their assets. Thus, when considering pedestrians or agents with decision-making capacity, two different situations may occur: equilibrium ($N_a \sim N_b$) or herding ($N_a \gg N_b$ or $N_a \ll N_b$).

The decision mechanism works as follows:

Each particle reaching the line of the door must choose between keeping and changing its state (from $a$ to $b$ or vice versa).
In each cycle of reinsertion, this decision is made only once when a pedestrian enters a rectangular area $A$ of size $L$ x 1 m, extending 50 cm to both sides of the door. For the deciding pedestrian $i$, the following fraction is calculated: 
\begin{equation}
\xi_i = \frac{n_s}{(n_s+n_d)} ,
 \label{fract_x}
\end{equation}
where $n_s$ is the number of agents in the same state as agent $i$ and $n_d$ is the number of agents in the other state (with an opposite desired walking direction). Note that $n_s + n_d$ is the total number of agents in the area $A$.
Then, the decision is made by choosing a random number ($\chi$) from a uniform distribution in the interval [0, 1] and comparing it with a sigmoid function
\begin{equation}
F(\xi)={ 1 \over 1+\exp \left(\frac{-(\xi-0.5)}{T}\right) } .
\label{sig}
\end{equation}
If $F(\xi_i) <  \chi$, then agent $i$ changes its state. Because of the shape of this function, it is more probable to change state if the particles are in a minority. In other words, if pedestrian $i$ is in the minority, it is more likely to join the state of the majority.

The parameter $T$ is treated as a behavioral parameter. When this parameter is low, the agents tend to show herding behavior, which means that they have a greater tendency to imitate the behavior of neighbors. On the contrary, when $T$ tends to infinity, the agents ignore the state of their neighbors. So, we call $T$ the ``individualistic'' parameter.

As $T$ goes to 0, $F(\xi)$ approaches to the step function, which describes a deterministic rather than probabilistic decision behavior (agents in the minority side will change their state and agents in the majority side will never change their state). In this extreme case, the number of particles of one type will saturate and this state will not be reverted.

An equilibrium between both populations is achieved if $T$ tends to $\infty $. Then, $F_{(\xi)}$ become $0.5$, which makes the decision totally random for each agent. No matter what the fraction of particles of one kind ($\xi$) is, the decision to change the state is made with probability $1/2$. In this case, the number of particles of each type fluctuates around the equilibrium value $N_a \sim N_b \sim  N_p/2$ (= 50\% of the population).

\section{Results}
In the following, we will explore the analogy between (a) the financial time series introduced in Sec. \ref{s21} and (b) the density of pedestrians around the door according to the models described in Sec. \ref{s22}. We will study both agents without and with decision-making capacity (see Secs. \ref{s221} and \ref{s222}). This will be done by analyzing the statistical properties of the return and related quantities defined in the next subsection. 
\subsection{Density versus Logarithm of the Price}
\label{}
Let $Y$ be any general time series. Then, we defined the return as
\begin{equation}
R_{Y}=\frac{dY}{dt}   .
\label{Ret}
\end{equation}
For a discrete time series, the discretely sampled return is
\begin{equation}
R_{Y}^k={Y(t_i+k)}-{Y(t_i)}   ,
\label{RetDisc}
\end{equation}
where $t_i$ indicates the discrete time steps and $k$ the number of time steps over which the return is computed.
In the particular case of an asset price ($S$), we take $Y=\log(S)$ where $S$ 
is the asset price, and $k=1$. Then Eq. (\ref{RetDisc}) becomes the well-known logarithmic return for financial time series ($R={\log(S(t_i+1))}-{\log(S(t_i))}$).

Now, for a general time series $Y$, we define the absolute value of the return by
\begin{equation}
|R_{Y}^k| = | {Y(t_i+k)}-{Y(t_i)} |
\label{AbsRetDisc}
\end{equation}
and the standardized absolute return (inspired by the standardized return \cite{Andersen2000, Virasoro2011}) by 
\begin{equation}
|{\hat R_{Y}^k} | = \frac{|R_{Y}^k |} {(\sum_{t_i=1}^{N_{T}-k}~ |R_{Y}^k|) / N_{T}}   ,
\label{NorAbsRetDisc}
\end{equation}
where $N_T$ is the total number of data points in the time series $Y$. Therefore the denominator is the arithmetic mean of the absolute return.

As stated above, for a price time series $S$, it is common to consider the logarithmic return by taking $Y ~ = ~ \log(S)$. In the case of the pedestrian system, the time series to be analyzed is the density, i.e.: $Y=\rho$. We calculate the density as the average density over three equidistant points on the door line, by using the $\kappa$-nearest neighbor algorithm ($\kappa$-NN) with $\kappa = 8$. This algorithm consists in measuring the distance ($d_{\kappa}$) to the $\kappa^{th}$ nearest neighbor from any point $(x_0,y_0)$, and so the density in that point can be approximated as $\rho_{(x_0,y_0)}=\frac{(\kappa-1)}{\pi {(d_k)}^2}$. 

All the stylized facts, for the time series from both systems, are observed for the return and related quantities defined above. The direct observables $\rho$ and $log(S)$ will not be compared quantitatively; however, it is worth noting that they have some differences and similarities depending on the scale.

For examples at large scale ($\sim10^4$ time steps):

- The pedestrian time series display an upper bound because the system can reach a maximum density; on the contrary, $log(S)$ is not upper bounded.

- Because of the dynamics of the pedestrian system and the fixed number of agents (of each kind $a$ and $b$), the density shows a periodicity (when decision is added, the periodicity is less defined). But financial prices do not show the same kind of periodicity, they could show trends at this scale.

\begin{figure}
\begin{center}
\centerline{ \includegraphics[scale=0.5]{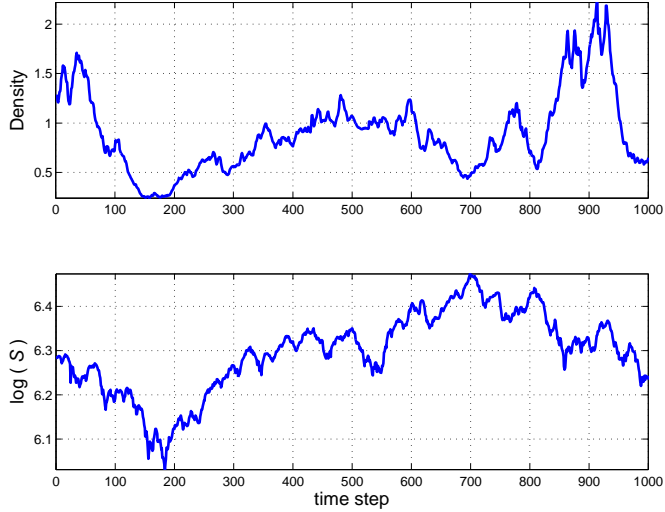}}
\caption{Qualitative comparison between arbitrary windows ($10^3$ time steps) of the original time series of both systems: density of pedestrians around the door (top) and NYSE100-1day (bottom)}
\label{Fig_OTS}
\end{center}
\end{figure}

Similarities can be seen at smaller scales. In fig. \ref{Fig_OTS}, a comparison is made between a price time series ($log(S)$) and the density time series of one realization of the pedestrian system, considering agents with decision-making capacity, at the scale of $10^3$ time steps.

\subsection{Automata Agents}
\label{sec:3:2}
In this subsection, we study the stylized facts of the pedestrian system of automata agents as described in Sec. \ref{s22}. These are compared with those emerging from financial markets.
The parameters characterizing the pedestrian system are: $L= 7$ m; $N_p= 60$ (30 pedestrians in state $a$ and 30 in state $b$ all the time). The total number $N_p= 60$  of particles used in our simulations 
should be compared with the number (approximately $600$) that would correspond to a random close packing.
Hence, the density of particles is comparable to a dense gas for which the effect of the
constriction is essential for the properties described below to emerge.
\subsubsection{Heavy Tails}
\label{}
\begin{figure}
\begin{center}
\centerline{ \includegraphics[scale=0.5]{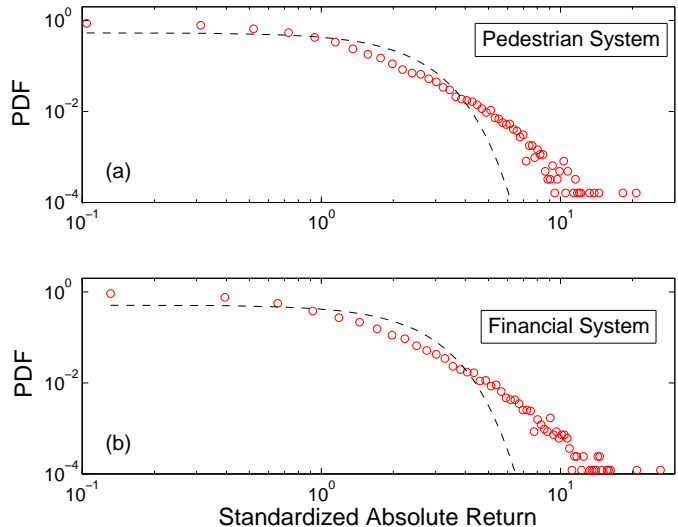}}
\caption{Probability density functions (PDF) of the standardized absolute returns, (a) of the time series $\rho$ for simulated automata pedestrians and (b) of the time series $\log(S)$ for the EURUSD-1 hour currency exchange data. The dashed lines indicate the closest Gaussian distribution as a guide to the eye to reveal the fat tails of both PDF.}
\label{Dist}
\end{center}
\end{figure}

An important characteristic of distribution functions of return of financial time series is that they exhibit fatter tails than a Gaussian distribution \cite{MalPisSor}. In order to make both time series comparable, we take the standardized absolute return. Figure \ref{Dist} shows both distributions, where the dotted line indicates the closest Gaussian distribution as reference.
It can be seen that both distributions exhibit fat tails and look very similar. 

\subsubsection{Aggregational Gaussianity}
\label{AgreGauss}
\begin{figure}
\begin{center}
\centerline{ \includegraphics[scale=0.5]{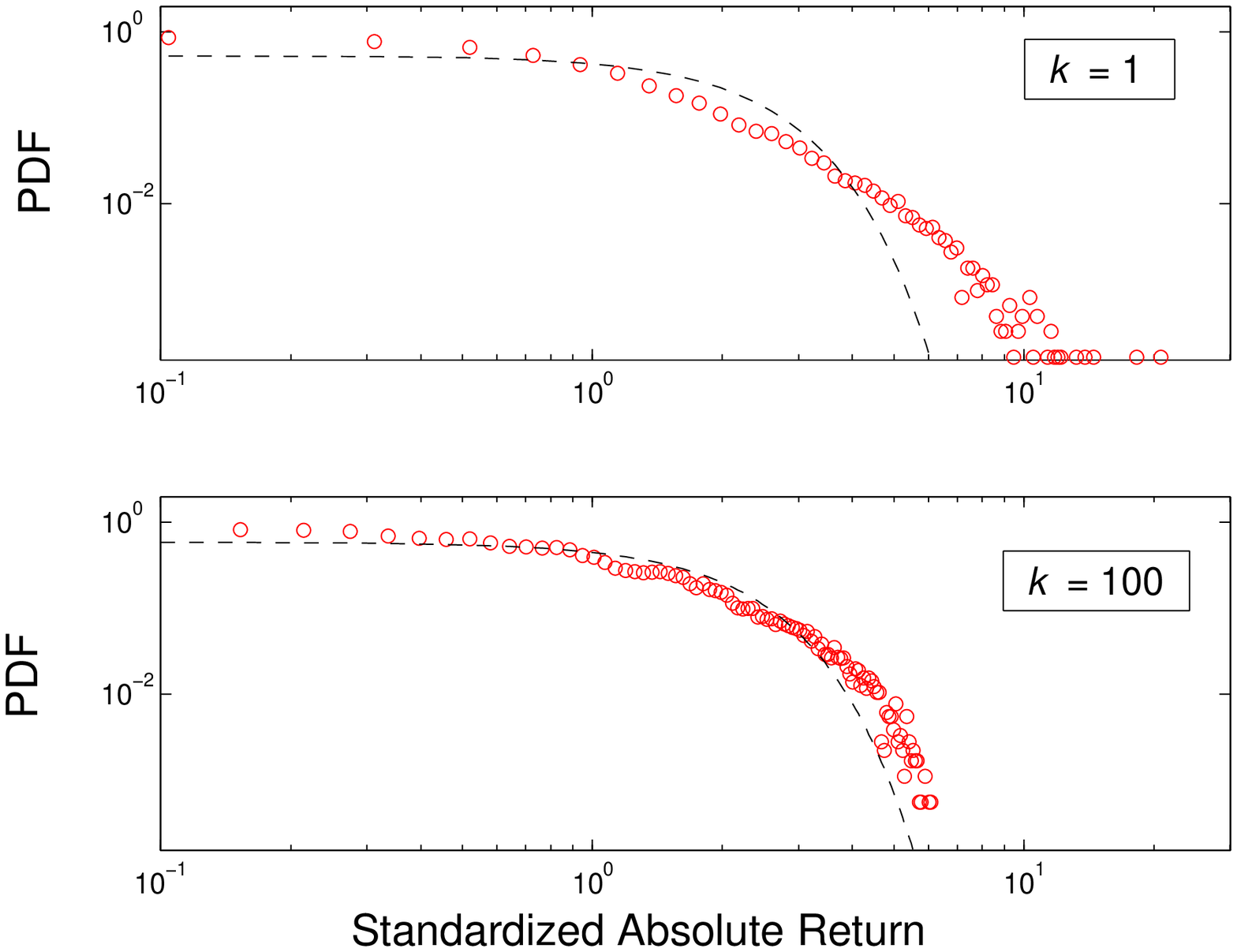}}
\caption{Probability density function of the standardized absolute return of $\rho$ with $k$ = 1 (top) and $k$=100 (bottom). The dashed line indicates the closest Gaussian distribution.}
\label{AgrAutomata}
\end{center}
\end{figure}

As seen in Eqs. (\ref{RetDisc}) - (\ref{NorAbsRetDisc}), for discrete time series, the return can be calculated over different periods of time, parameterized by the number of time steps ($k=1,~2,~...$).
It is known for financial data that, as the number of time steps $k$ increases, the distribution of the return converges against a Gaussian distribution \cite{Cont2001,BouchaudPotters}. Figure \ref{AgrAutomata} shows this tendency for $|{\hat R_{\rho}^k}|$ for the pedestrian system.
\subsubsection{Autocorrelation of Return and Volatility}
\label{}
\begin{figure}
\begin{center}
\centerline{ \includegraphics[scale=0.5]{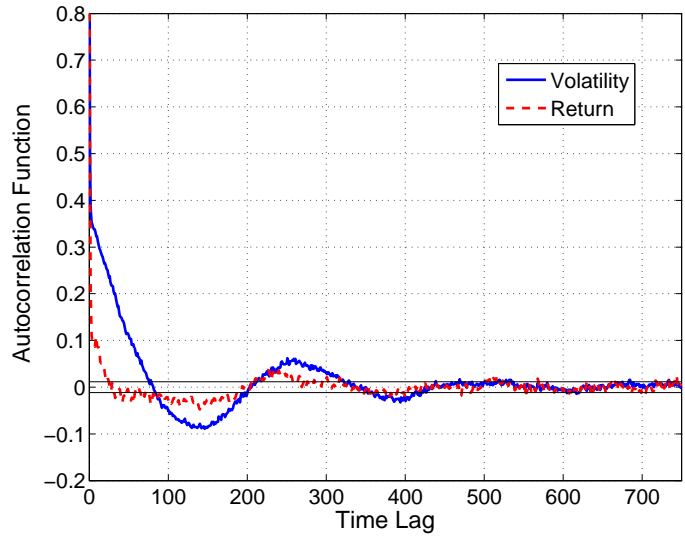}}
\caption{Autocorrelation functions of the return and absolute return of the density time series for automata pedestrian system.}
\label{autocorr}
\end{center}
\end{figure}

Asset returns show no linear autocorrelation, except for very small time scales. However, the volatility displays a positive autocorrelation. This indicates that big price fluctuations are often followed by big price fluctuations, a fact which is known as ``volatility clustering" \cite{Mandelbrot,Cont2001,Calvet-Fischer,Cont07}.
Taking the absolute return as a measure of the volatility, we compare the autocorrelation function of the return and the absolute return for the pedestrian time series (Fig.\ref{autocorr}).

The plot reveals that the system of automata agents presents an autocorrelation of absolute returns that is greater than the autocorrelation of returns. Although this phenomenon is visible in the pedestrian system, the decay of the autocorrelation function of absolute returns is much slower in financial time series. This stylized property will become more similar in the two time series when decision-making capacity is added to agents (see Sec. \ref{autocrr_decision}).

\subsubsection{Scaling of the Peaks of the Distribution of Returns}
\label{}
\begin{figure}
\begin{center}
\centerline{ \includegraphics[scale=0.5]{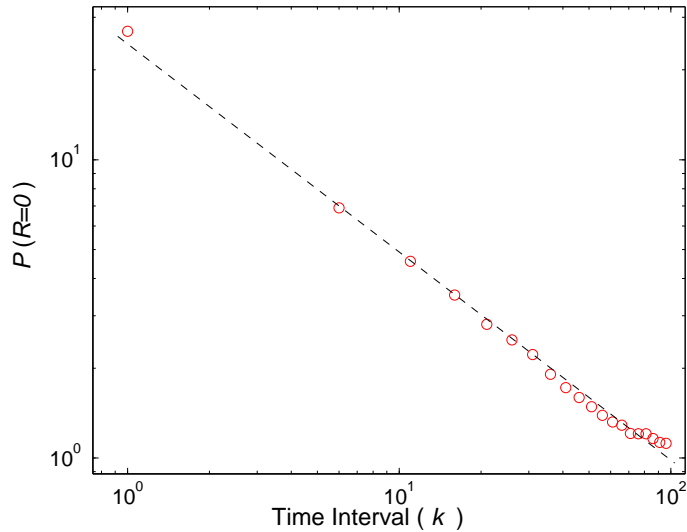}}
\caption{Maxima of the distributions of returns $R_{\rho}^k$ versus the number $k$ of time steps.}
\label{autScalingPeaks}
\end{center}
\end{figure}

As in Sec. \ref{AgreGauss}, we calculate here the return ($R_{\rho}^k$ and $R_{\log(S)}^k$) for different time intervals $k$. For the pedestrian system and FX markets, the probability distribution of returns is symmetric and has a peak at zero return. It was shown in Ref. \cite{Mantegna1995} that the S\&P500 index exhibits a power-law scaling behavior when the probability of zero return ($P_{(R=0)}$) is plotted against the time interval $k$. In our case, we approximate the probability of return $R^k=0$ by computing the kernel density estimator \cite{Botev2010}.
The values of $k$ considered were $k=1, 6, 11, ...,101$. Figure \ref{autScalingPeaks} displays a log-log plot of the probability ($P_{(R=0)}$) versus $k$.
The figure shows that the scaling property exists also in the pedestrian system. Furthermore, the value of the power-law exponent is $\alpha= -0.70$, very similar to the one observed for financial data: EURUSD-1min ($\alpha=-0.71$), EURUSD-10min ($\alpha=-0.77$), CHFUSD-1hour ($\alpha=-0.67$). Moreover, Ref. 
\cite{Mantegna1995} reports $\alpha=-0.71$ for S\&P500.
\subsubsection{Multifractality}
\label{}

\begin{figure}
\begin{center}
\centerline{ \includegraphics[scale=0.5]{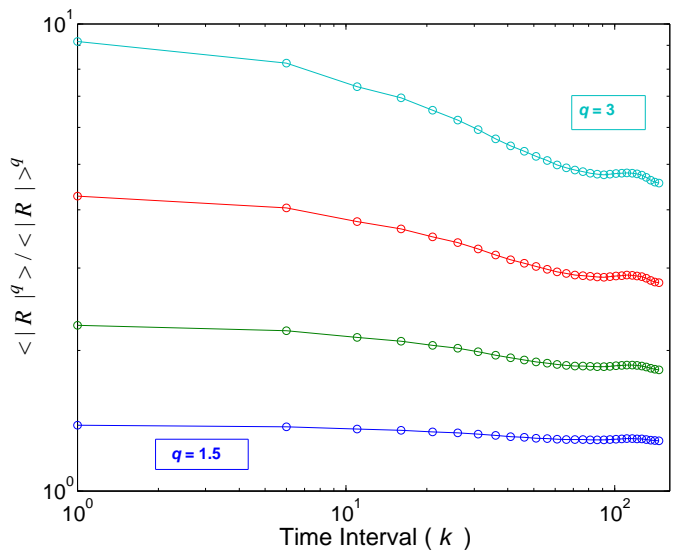}}
\caption{Ratio $<|R_{(\rho)}^k|^q>/<|R_{(\rho)}^k|>^q$ versus number $k$ of steps over which the return is calculated, for $q$ = 1.5, 2, 2.5 and 3. }
\label{Multfrac}
\end{center}
\end{figure}

Multifractality can be tested by examining the ratio $<|R^k|^q>/<|R^k|>^q$ for returns calculated with different time steps $k$. This ratio is constant for a simple fractal, 
but not for a multifractal \cite{Calvet-Fischer}. 
Figure \ref{Multfrac} shows this plot for the pedestrian system considering $q$ = 1.5, 2, 2.5 and 3.

If we take the slope of the higher curve ($q=3$) as a measure of multifractality ($-0.16$), it is very similar to the ones obtained for financial time series (ranging from $-0.13$ to $-0.18$).
\subsubsection{Hurst Exponent}
\label{}
Self-similarity of the signal is an important feature and has been largely studied for financial systems \cite{Mandelscaling,LaSpada2008,Lux1999}.
In this subsection, we calculate the Hurst exponent ($H$) of the absolute return time series ($|R_{Y}|$). This study was also performed in Ref. \cite{LaSpada2008} by analyzing four major stocks from the London Stock Exchange.

We calculate $H$ using the Detrended Fluctuation Analysis (DFA) \cite{Peng1995,Lux1999}. The value of $H$ obtained for the pedestrian system is $H = 0.88$, which is similar to the ones obtained for DJI-1day ($H = 0.86$) and NYSE100-1day ($H = 0.85$). Furthermore, the results reported in Ref. \cite{LaSpada2008} are also similar, ranging from $H = 0.80$ to $H = 0.86$. The other financial time series analyzed in the present work display values of $H \geq 0.65$. 
In all cases, the results show that self-similarity is present, and the Hurst exponent lies in the region $0.5 < H < 1$, which corresponds to correlated noise, indicating long-term memory of the absolute return.

\subsection{Pedestrians with Decision-Making Capacity}
\label{}
In previous sections, we analyzed the statistical properties of the pedestrian system for automata agents; here we will study the properties when we provide simulated pedestrians with a decision-making capacity. As explained in \ref{s222}, this allows us to consider herding effects, so that the system may develop a behavior similar to financial bubbles and crashes.

The parameters of the system for this behavior are the same as before,  $L = 7$ m and $N_p = 60$. At time $t=0$, we assume again $N_a = N_b = N_p / 2 = 30$,  but as the system evolves, $N_a$ and $N_b$ will change. 
\subsubsection{Variations in the Decision Parameter}
\label{var_beta}
\begin{figure}
\begin{center}
\centerline{ \includegraphics[scale=0.5]{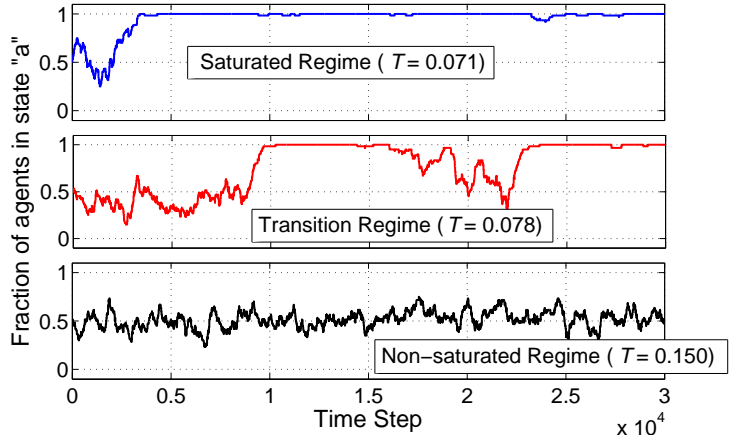}}
\caption{Comparison of the fraction $\frac{N_a}{N_p}$ for 3 realizations for different values of the individualistic parameter $T$. }
\label{Pop}
\end{center}
\end{figure}

A relevant observable is the fraction of particles in a certain state, for example, let us take the state $a$. Figure \ref{Pop} shows the evolution of the fraction $N_a / N_p$ for three different values of $T$. 

The system can be found in three different regimes.

- Saturated regime: For low values of the individualistic parameter ($T \lesssim 0.07),$ the system saturates rapidly, indicating that all simulated pedestrians have the same state and thus a unidirectional flow is established.

- Non-saturated regime: For $T \gtrsim 0.09$,  the system oscillates around $N_a / N_p \sim 1 / 2$. In this regime, the system behaves very similar to automata agents (Sec. \ref{sec:3:2}).

- Transition regime: For $0.07 \lesssim T \lesssim 0.09$, the behaviors of the saturated and non-saturated regimes may be combined in the same realization (see middle panel of Fig. \ref{Pop}).

\begin{figure}
\begin{center}
\centerline{ \includegraphics[scale=0.5]{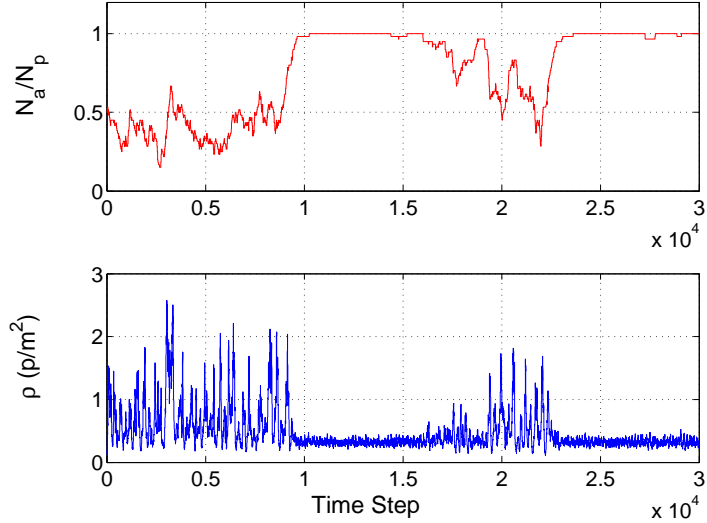}}
\caption{Fraction of the population $\frac{N_a}{N_p}$ for $T=0.078$ (top) and related pedestrian density time series (bottom). }
\label{Sat_Ori}
\end{center}
\end{figure}

When the system remains saturated, all pedestrians end up acting in the same way. This situation might be analogous to stock market panic generating a financial crash. In fact, saturation periods are correlated with low values of the density time series (0.2 p/m$^2$  $\lesssim \rho \lesssim$ 0.5 p/m$^2$), as shown in Fig. \ref{Sat_Ori}. This feature further justifies the analogy between the pedestrian density ($\rho$) and the logarithm of the price ($\log(S)$). 

We observe that, if only periods of unidirectional flow occur, all stylized facts disappear. Therefore having two groups of agents with opposite goals (counterflow) is of fundamental importance for the emergence of the stylized facts studied in this paper.

\subsubsection{Autocorrelation of Return and Volatility}
\label{autocrr_decision}
\begin{figure}
\begin{center}
\centerline{ \includegraphics[scale=0.5]{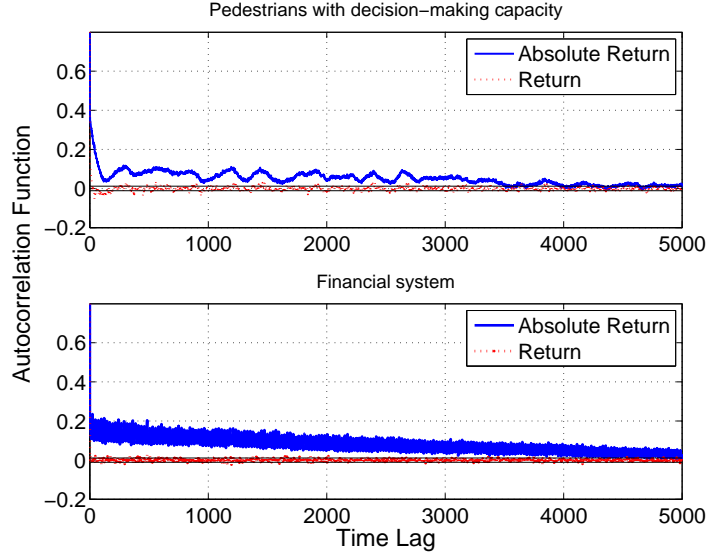}}
\caption{Autocorrelation functions of the return and absolute return for pedestrians with $T = 0.078$ (top) and for EURUSD-1hour financial data (bottom).}
\label{Autocorr_beta}
\end{center}
\end{figure}

Consider the non-saturated regime ($T \gtrsim 0.09$) in which the number of pedestrians belonging to each group ($N_a$ or $N_b$) fluctuates around the equilibrium value $N_p / 2$.
In this situation, the autocorrelations are very similar to the one observed in Fig. \ref{autocorr} for automata agents where the number of pedestrians in each state is fixed ($N_a = N_b = N_p / 2$).
However, for values of $T$ in the transition zone ($0.07 \lesssim T \lesssim 0.09$), the time period above which the autocorrelation of the absolute return becomes negligible is much larger, in accordance with financial time series. 
In Fig. \ref{Autocorr_beta}, it can be seen that both systems have a similar autocorrelation function of returns and absolute returns. 

This result for the pedestrian system holds, in general, for $T$ in the transition regime. The length of the decay of the autocorrelation depends on the coexistence of periods of saturation and non-saturation of the population in the same realization, which is more likely to occur near $T \sim 0.08$.

It is important to stress that, in the saturated regime (corresponding to the unidirectional flow), all stylized facts (and in particular the autocorrelation of absolute return) disappear. While in the non-saturated regime with permanent counterflow ($T \gtrsim 0.09$), the autocorrelation of absolute return is weak (see Fig. \ref{autocorr}), independently of whether decision-making capacity is considered or not. Thus, it is interesting to discover that the significant autocorrelation of absolute returns can be understood by the combination of the two different regimes (the saturated and non-saturated one). In other words, this stylized fact appears when the pedestrian system is in the transition area between the saturated `herding' regime and the non-saturated `equilibrium' regime. Such behavior is also found in other systems, such as generalized Ising models
for which the excess volatility that is typical of real financial markets is obtained in the transition
region of the underlying Ising critical point \cite{HarrasTessoneSor}.

\section{Conclusions}
In this work, we have shown that a pedestrian counterflow system with bottleneck exhibits several stylized facts that are characteristic for financial systems, if the pedestrian density observed in the conflict zone (around the door) is compared to the logarithm of the price of financial assets. Hence, the simulated pedestrian system may become an excellent tool to get deeper knowledge of what causes stylized fact observed in financial systems.

We find that having two groups of agents with opposite interests (counterflow) is an important precondition to reproduce the stylized facts. Already when these two groups have a constant number of agents during the simulations (automata agents), we find the following stylized facts: heavy tails of the distribution of returns, slow decay of the autocorrelation of absolute return, aggregational Gaussianity, scaling of the peaks of the distribution of returns, multifractality and self-similarity.
Having agents with adaptive behavior (i.e., changing their state by making decisions) is only relevant to explain the slowness of decay of the autocorrelation of absolute returns. This is due to the occurrence of herding behavior, when agents can change their state, making their decisions depending on the state of the neighbors. 

The tendency of agents to mimic other agents is controlled by the parameter $T$.  Depending on its value, the system can be in three regimes: (a) a saturated regime ($T \lesssim  0.07$), where all the agents are in the same state; (b) a non-saturated regime ($ T \gtrsim  0.09$), where the population of both classes of agents are in equilibrium (which is similar to the behavior of automata agents with a fix number of agents in each class) and (c) a transition regime ($0.07 \lesssim  T \lesssim  0.09$) in which the saturated and the non-saturated regimes alternate during intermittent time periods. In the transition regime, the decay of the autocorrelation of absolute return is much slower than the decay of the autocorrelation of the return, matching very well this interesting stylized fact observed in financial time series.  

The properties reported here, obtained for a system with $60$ particles, do not 
change appreciably with varying this number within factors of $2$.
However, we should stress that the characteristics of the density fluctuations
that are so similar to financial price fluctuations depend on the system 
not being too large. In other words, the stylized facts disappear in the thermodynamic limit
and are thus intrinsically ``finite-size effects''.
This should not be taken necessarily as a drawback, since there is evidence
that the price dynamics of any given financial asset is typically driven
by no more than about one hundred investors. We refer to Refs.~\cite{Corcos,Sortaka}
for reviews of the finite-size effects in various models of financial price dynamics.

In conclusion, the statistical analogies between the pedestrian counterflow problem and  financial time series suggest that studying pedestrian counterflow systems in the presence
of constraints may help to gain a better understanding of the mechanisms underlying stylized facts of financial markets. This opens the road to the understanding and characterization of certain abnormal market regimes, 
in analogy with the corresponding pedestrian systems. 
For instance, when the bottleneck is narrower (opening $L<4$ m), permanent
blockage occur due to the soft attraction, which nucleate clusters around and
within the constriction. In finance, this corresponds to the no-trade situation
occurring when the liquidity vanishes. It would be interesting to study 
the properties of the density fluctuation close to this jamming transition
in parallel to the corresponding situation in financial markets.
Future works will also investigate the different regimes of herding behavior
within the proposed analogy. As the density of particles is varied from dilute gas, to
liquid and glass, the effect of the bottleneck can be studied systematically
with respect to its impact on the density fluctuations in its neighborhood.
Similarly to the physics of lubrication and/or of wetting, novel critical behavior
and transitions can be expected to translate into an interesting classification
of fluctuations in systems characterized by balanced flows with bottlenecks.

\begin{acknowledgments}

D.R.P. is grateful for partial financial support by ETH Z\"{u}rich and by CONICET (via the project P.I.P. - 0304).\\

\end{acknowledgments}

\end{document}